\documentclass[aip,jcp,reprint,amsmath,amssymb]{revtex4-1}
\usepackage{graphicx}
\usepackage{dcolumn}
\usepackage{bm}
\usepackage[utf8]{inputenc}
\usepackage[T1]{fontenc}
\usepackage{mathptmx}
\usepackage{etoolbox,float}
\usepackage{booktabs}

\makeatletter
\def\@email#1#2{%
 \endgroup
 \patchcmd{\titleblock@produce}
  {\frontmatter@RRAPformat}
  {\frontmatter@RRAPformat{\produce@RRAP{*#1\href{mailto:#2}{#2}}}\frontmatter@RRAPformat}
  {}{}
}%
\makeatother

\begin{document}
\preprint{}

\title{Vacancy diffusion on a brominated Si(100) surface: Critical effect of the dangling bond charge state}

\author{T. V. Pavlova}
 \email{pavlova@kapella.gpi.ru}
\affiliation{Prokhorov General Physics Institute of the Russian Academy of Sciences, Vavilov str. 38, 119991 Moscow, Russia}

\author{V. M. Shevlyuga}
\affiliation{Prokhorov General Physics Institute of the Russian Academy of Sciences, Vavilov str. 38, 119991 Moscow, Russia}
\affiliation{HSE University, Myasnitskaya str. 20, 101000 Moscow, Russia}


\begin{abstract}

Silicon dangling bonds (DBs) on an adsorbate-covered Si(100) surface can be created in a scanning tunneling microscope (STM) with high precision required for a number of applications. However, vacancies containing DBs can diffuse, disrupting precisely created structures. In this work, we study the diffusion of Br vacancies on a Si(100)-2$\times$1-Br surface in an STM under typical imaging conditions. In agreement with previous work, Br vacancies diffuse at a positive sample bias voltage. Here, we demonstrated that only vacancies containing a positively charged DB hop across the two atoms of a single Si dimer, while vacancies containing neutral and negatively charged DBs do not. Calculations based on the density functional theory confirmed that positively charged Br (and Cl) vacancies have a minimum activation barrier. We propose that diffusion operates by both one-electron and two-electron mechanisms depending on the applied voltage. Our results show that the DB charge has a critical effect on the vacancy diffusion. This effect should be taken into account when imaging surface structures with charged DBs, as well as when studying the diffusion of other atoms and molecules on the Si(100) surface with vacancies in an adsorbate layer.

\end{abstract}

\maketitle

\section{Introduction}
Vacancies in the adsorbate layer on the Si(100) surface contain silicon dangling bonds (DBs), which become charged by gaining or donating an electron. Chemically active DBs can be employed in selective chemistry, in particular, for doping silicon with near-atomic precision \cite{2003Schofield, 2020Stock}. In addition, DBs could be assembled on Si(100) to construct tunnel coupled structures \cite{2011Pitters}, which can be used as artificial quantum systems \cite{2013Schofield, 2018Wyrick} or atomic scale logic devices \cite{2018Huff}. For these applications, single vacancies in the adsorbate layer are created in a controlled manner in a scanning tunneling microscope (STM) with high accuracy. However, vacancies can diffuse over the surface, so it is important to know what influences their diffusion to avoid uncontrolled movement of DBs.

Vacancies on Si(100)-2$\times$1-H serve as a model system for studying diffusion over semiconductor surfaces. Diffusion of H vacancies was investigated both experimentally and theoretically in a substantial number of works at various temperatures, sample voltages, and currents (see review \cite{2013Durr} and references therein). Although the influence of the DB charge on the vacancies diffusion has not yet been systematically studied, there are several indications that the charge of a vacancy affects the diffusion process. In particular, it was theoretically found that the transfer of an electron from a neutral vacancy to a neighboring one reduces the diffusion activation energy \cite{2010Wieferink}. In Ref. \cite{2000Stokbro}, it was shown that the holes injection into surface states creates a local positive charge, which affects the vacancy diffusion on the Si(100)-2$\times$1-H surface. It was also noted in other experimental works that diffusing vacancies contained positively charged DBs \cite{2002Durr, 2007Schwalb}.

The present study is devoted to the vacancy diffusion on a Si(100)-2$\times$1-Br surface, which has the same structure as Si(100)-2$\times$1-H. This is a continuation of our previous work \cite{2022Pavlova}, in which we demonstrated the possibility of controlled creation of single vacancies and changing their charge state on a brominated Si(100) surface. To the best of our knowledge, the vacancies diffusion on Si(100)-2$\times$1-Br was previously considered only in one work \cite{2002Nakayama}. Based on the dependence of the hopping rate on current and voltage, a diffusion mechanism was proposed as the transfer of an electron to the antibonding Si-Br orbital. In this experiment, diffusion over the surface was studied without division into three pathways (across the two atoms of a single Si dimer, along to the surface dimer rows, and across the dimer rows). Importantly, it was shown that vacancies move only at a positive voltage \cite{2002Nakayama}. However, the influence of the DB charge on vacancy diffusion was not taken into account.

In this work, we studied the effect of the DB charge on the vacancy diffusion on a brominated Si(100) surface. We considered vacancy diffusion across the two atoms of a single dimer, since such a process is the fastest \cite{2013Durr} and, therefore, has a main impact on the DB displacement. Scanning tunneling microscopy of the Si(100)-2$\times$1-Br surface at a positive voltage clearly indicate that only vacancies containing positively charged DB (DB$^+$) diffuse, while vacancies containing neutral DB (DB$^0$) or negatively charged DB (DB$^-$) do not diffuse. Calculations based on the density functional theory (DFT) confirm the lowest activation energy for vacancy diffusion with DB$^+$. If the voltage polarity is reversed to negative, the vacancies that have diffused under positive voltage no longer move and, therefore, they can be imaged in an STM without the risk of destroying the assembled DB structure.

\section{Experimental and computational details}

The experiments were carried out in an ultra-high vacuum (UHV) system with a base pressure of 5$\times$10$^{-11}$\,Torr. The STM measurements were performed with GPI CRYO (SigmaScan Ltd.) operated at 77\,K. The Si(100) samples were prepared by outgassing the wafer at 870\,K for several days in UHV followed by flash-annealing at 1470\,K. Subsequently, the samples were passivated with Br (Cl) at a partial Br$_2$ (Cl$_2$) pressure of 10$^{-8}$\,Torr for 100--200\,s at a temperature of 370--420\,K. To study the diffusion of Br vacancies, p-type Si(100) samples (B-doped, 1\,$\Omega$\,cm) were used. STM images of the diffusion of a Cl vacancy were recorded on an n-type Si(100) sample (P-doped, 0.1\,$\Omega$\,cm). We used both electrochemically etched polycrystalline W tips and mechanically cut Pt-Rh tips. The voltage ($U_s$) was applied to the sample. All STM images were processed using the WSXM software \cite{WSXM}.

The spin-polarized DFT calculations were performed with the Perdew-Burke-Ernzerhof (PBE) functional \cite{1996Perdew} as implemented in the Vienna \textit{ab initio} simulation package (VASP) \cite{1996Kresse, 1999Kresse}. The kinetic-energy cutoff of the plane wave basis was set to 400\,eV. To account van der Waals correction we used DFT-D2 method developed by Grimme \cite{2006Grimme}. The Si(100)-2$\times$1 surface was modeled as a 16-layer slab with a 6$\times$6 supercell and a 15\,{\AA} vacuum gap. Br (Cl) atoms were placed on the top surface to form the Si(100)-2$\times$1-Br (-Cl) structure, whereas the bottom Si surface was passivated with hydrogen. The bottom two Si layers were frozen at their bulk positions, while the coordinates of other atoms were fully relaxed until the residual forces were smaller than 0.01\,eV/\,{\AA}. Brillouin zone integrations were done using a 4$\times$4$\times$1 k-point grid. To simulate positive or negative charge states, an electron was removed or added to the supercell, respectively. For bonding analysis, the crystal orbital Hamilton population (COHP) \cite{1993Dronskowski} was evaluated using LOBSTER \cite{2011Deringer, 2013Maintz, 2016Maintz}. For COHP calculations, the 3$\times$4$\times$16 cell was used with the 4$\times$3$\times$1 k-points grid. STM images were calculated within the Tersoff--Hamann approximation \cite{1985Tersoff}. The voltage in simulated STM images is indicated relative to the valence-band maximum.

The activation barriers were calculated by using the nudged elastic band (NEB) method \cite{1998NEB}. In this case, the criterion of convergence for residual forces was reduced to 0.03\,eV/\,{\AA}. We used five images, including the two end points. For an adatom diffusing across the two atoms of a single dimer, the coordinate along the reaction path was fixed to prevent image movement along this path. For a positively charged vacancy, the barrier was additionally refined with five images between the initial and bridge configurations.

\section{Results}

\subsection{Experimental results}

\begin{figure}[h!]
 \begin{center}
 \includegraphics[width=\linewidth]{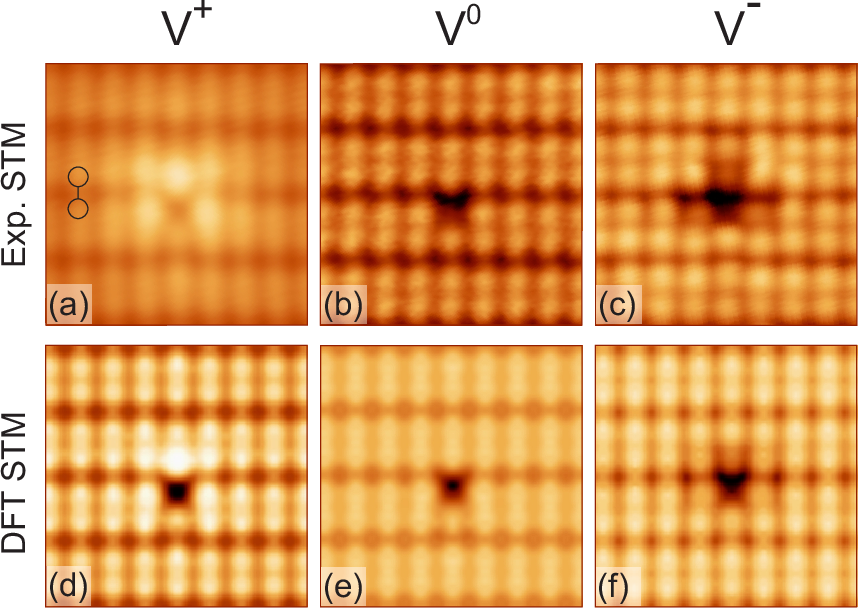}
\caption{\label{figSTM} Experimental (a--c) and simulated (d--f) empty state STM images of the V$^+$, V$^0$, and V$^-$ on the Si(100)-2$\times$1-Br surface. Experimental images were obtained at I$_t$ = 1.7\,nA; $U_s =+2.8$\,V (a), I$_t$ = 1.7\,nA; $+2.5$\,V (b), I$_t$ = 2.0\,nA; $+2.0$\,V (c). Theoretical images were simulated at $U_s =+2.8$\,V (d), $U_s =+2.5$\,V (e), and $U_s =+2.0$\,V (f). Silicon dimer is marked by a dumbbell in (a). }
\end{center}
\end{figure}

The Si(100)-2$\times$1-Br surface consists of rows of Si dimers, in which each Si atom is terminated by a Br atom. In place of the Br vacancy, a Si dangling bond is formed, which has three charge states. We assign the same charge to the vacancy as the charge state of the DB, i.e. a vacancy with the DB$^+$, DB$^0$, and DB$^-$ will be denoted by the V$^+$, V$^0$, and V$^-$, respectively. In an empty state STM image, the neutral vacancy V$^0$ looks like a depression, while in the case of the V$^+$ (V$^-$) vacancy a depression is surrounded by a bright (dark) halo  (Fig.~\ref{figSTM}). The bright and dark halo in an empty state STM image is caused by the double unoccupied DB$^+$ and occupied DB$^-$ state, respectively. The halo is well reproduced in the simulated STM images (Fig.~\ref{figSTM}) and is consistent with STM images of charged vacancies on a hydrogenated \cite{2015Labidi, 2013Schofield} and chlorinated \cite{2022Pavlova} Si(100) surfaces.

\begin{figure}[h]
 \includegraphics[width=\linewidth]{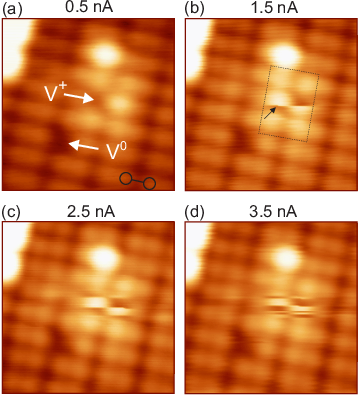}
\caption{\label{Br_diff} STM images (3$\times$3\,nm$^2$, $U_s =+2.5$\,V) of the hopping motion of the positively charged vacancy V$^+$ on Si(100)-2$\times$1-Br. The tunneling current is indicated above each STM frame. The neutral vacancy V$^0$  does not hop at any current. The recording time of each scan was 13 sec, the slow scan direction proceeded from bottom to top. An atomic step and a point defect are present in the upper part of the STM images. Silicon dimer is marked by a dumbbell in (a). The switching event is indicated by the arrow in (b). The surface area used to calculate the hopping rate is shown by the rectangle in (b).}
\end{figure}

Figure~\ref{Br_diff}a shows the STM image of Si(100)-2$\times$1-Br with two bromine vacancies, V$^+$ and V$^0$, which can be observed close together (see the supplementary material for additional STM images). Scanning with a positive voltage bias leads to the V$^+$ hopping across the two atoms of a Si dimer, which is visualized as a shift of the vacancy position between two successive line scans (Figs.~\ref{Br_diff}b--d). When scanning the V$^0$ and V$^-$ at currents 0.5--4.5\,nA and positive voltages 1.7--4.5\,V, we did not observe the same switching events (shifts in the line scan), as shown in Fig.~\ref{Br_diff} for the V$^+$. Nevertheless, we have very rarely observed a change in the position of the V$^0$ and V$^-$ in sequentially recorded STM frames (see the supplementary material for STM images of the V$^0$ and V$^-$ diffusion). Thus, only the V$^+$ hops at a positive voltage, and the frequency of its hops increases with the increase of the tunneling current $I_t$ (Fig.~\ref{Br_diff}). Note that we did not observe longer range V$^+$ hopping. When we reversed the voltage polarity to negative, we did not detect the hopping motion of vacancies that were positively charged at a positive voltage bias.

To investigate the mechanism of vacancy diffusion, we measured the dependence of the hopping rate on the sample voltage and tunnel current. The hopping rate ($R$) was calculated as the number of switching events divided by the time when the STM tip was over the surface area with a vacancy. This surface area with a vacancy was chosen to be three dimers in size (the rectangle in Fig.~\ref{Br_diff}b), since the shift between two line scans was observed mainly in this area. \footnote{The surface area with a vacancy of three dimers in size is about 0.3\,nm$^2$, so if a 3$\times$3\,nm$^2$ frame was recorded within 13 seconds (Fig.~\ref{Br_diff}), then this area was recorded within about 0.5 seconds. Then, for example, the hopping rate of 30\,$s^{-1}$ corresponds to about 15 vacancy jumps.} Nonlocal diffusion was also reported for hydrogen and chlorine \cite{2001Quaade, 2010Bellec}. To obtain each point on the graphs, we averaged $R$ over at least ten values, so the experimental data for one vacancy were collected over several hours. If during this time the tip state changed (for example, a secondary imaging feature appears), the collection of experimental data for a given vacancy was terminated. For this reason, we were unable to measure $R$ dependences both on current and voltage for all vacancies. Vacancies V$^+$1 and V$^+$2, as well as V$^+$4 and V$^+$5 were located on the same STM frame; therefore, the tip state was the same during their scanning. Vacancies V$^+$3, V$^+$4, and V$^+$5 were scanned with W tips; for the rest, Pt-Rh tips were used. We did not observed the influence of the tip material on the hopping rate.

\begin{figure}[t!]
 \includegraphics[width=\linewidth]{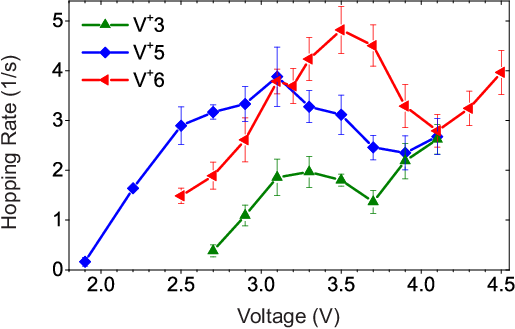}
\caption{\label{RU} Hopping rate of three different V$^+$s on Si(100)-2$\times$1-Br as a function of the voltage at $I_t = 1$\,nA. The lines are guides to the eye. The statistical error based on the number of hopping events is indicated by error bars.}
\end{figure}

Figure~\ref{RU} shows the hopping rate dependence on the sample voltage for three vacancies. $R (U_s)$ graphs for other three vacancies are given in the supplementary material. For all vacancies, the hopping rate has a broad peak in the range of 2--4\,V with a threshold at 2.0--2.5\,V and a maximum at 3.0--3.5\,V.

\begin{figure}[t!]
 \includegraphics[width=0.95\linewidth]{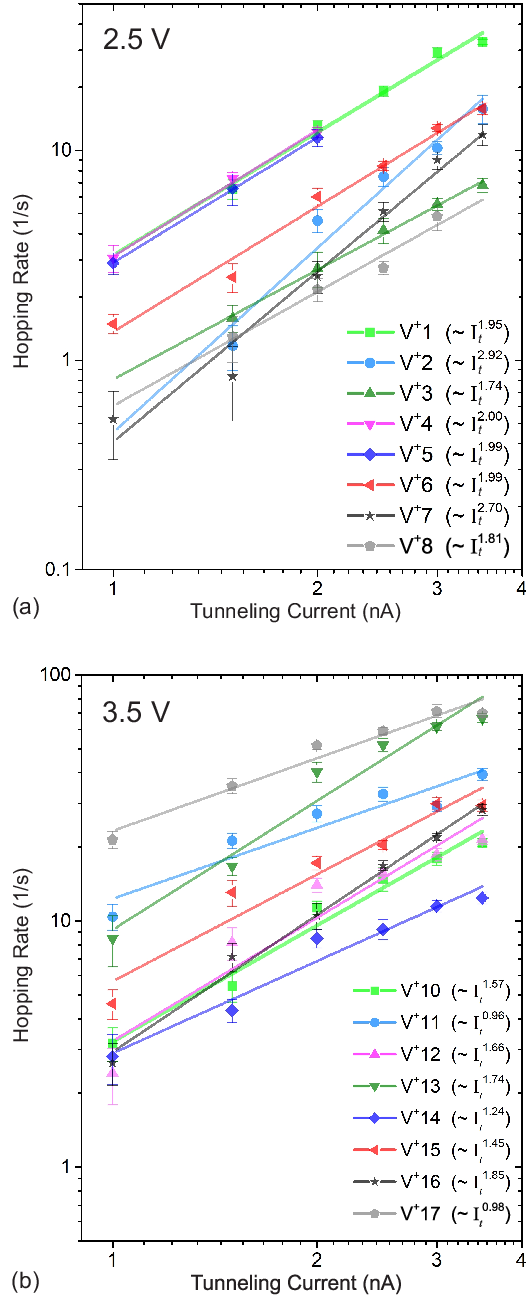}
\caption{\label{RI} Hopping rate of different V$^+$s on the Si(100)-2$\times$1-Br surface as a function of the tunneling current in the range from 1.0 to 3.5\,nA at $U_s =+2.5$\,V (a) and $+3.5$\,V (b). The lines fit the points by a power law $R \propto I_t^N$.  The statistical error based on the number of hopping events is indicated by error bars.}
\end{figure}

The hopping rate as a function of tunneling current was measured both at the threshold voltage of 2.5\,V (Fig.~\ref{RI}a) and at a voltage close to the maximum of the peak, 3.5\,V (Fig.~\ref{RI}b). Assuming $R = 0$ at $I_t = 0$, the hopping rate is best fitted by a power function $I_t^{N}$ with an exponent $N$ close to 2 at 2.5\,V and close to 1.5 at 3.5\,V. If we do not assume that $R$ vanishes at $I_t = 0$, then $R(I_t)$ is best fitted by a linear function. However, fitting the hopping rate with a linear function leads to $R>0$ or $R<0$ for different vacancies at $I_t = 0$. This behavior at zero current is difficult to explain, even if we take into account the thermal diffusion, because in this case $R$ would be positive for all vacancies at $I_t = 0$. Note that the assumption $R = 0$ at $I_t = 0$ is consistent with other studies of vacancies diffusion over the Br- \cite{2002Nakayama}, Cl- \cite{2003Nakamura}, and H-terminated \cite{2000Stokbro, 2001Quaade} Si(100) surfaces.

\begin{figure}[h]
 \includegraphics[width=\linewidth]{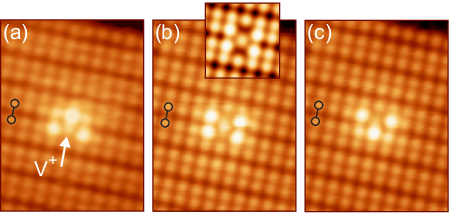}
\caption{\label{Cl_dif} Diffusion of a positively charged vacancy on the Si(100)-2$\times$1-Cl surface. STM images (3.8$\times$5.0\,nm$^2$, U$_s =+3.5$\,V, I$_t$ = 2.0\,nA) were recorded sequentially with a time interval of one minute. The vacancy in the lower position of the Si dimer (a) passes through the bridge configuration (b) to the upper position (c). Silicon dimer is marked by a dumbbell. The inset to (b) shows the simulated STM image of Cl in the bridge configuration at U$_s =+3.0$\,V.}
\end{figure}

On the Si(100)-2$\times$1-Cl surface, which has the same structure as Si(100)-2$\times$1-Br, we did not observe vacancy hopping when scanning at currents 0.5--3.5\,nA and positive voltages 1--5\,V. Nevertheless, we have rarely observed the displacement of a vacancy between sequentially recorded frames (Fig.~\ref{Cl_dif}). As in the case of bromine, a bright halo around the vacancy indicates that it contains the DB$^+$ \cite{2022Pavlova}. However, unlike bromine, chlorine is stable enough in the bridge configuration to be imaged in an STM (Fig.~\ref{Cl_dif}b) at 77\,K. These experimental data confirm that the atom passes through the bridge position when the vacancy moves \cite{1993Boland}, and that the V$^+$ moves much more slowly over a chlorinated Si(100) surface than over a brominated one \cite{2002Nakayama}.

\subsection{Computational results}

We calculated the activation barriers for intradimer diffusion of Br and Cl vacancies in each of the charge states. Note that the geometric and electronic structures for the DB$^+$, DB$^0$, and DB$^-$ on chlorinated and brominated surfaces are very similar \cite{2022Pavlova}. According to our calculations, the ferromagnetic configuration has a lower energy than the antiferromagnetic one for both the V$^+$ and V$^-$. Table~\ref{NEB} lists activation barriers for three charge states for Br and Cl vacancies. All energy barriers on the brominated surface are lower than on the chlorinated one. Remarkable, the activation barrier for the V$^+$ is much lower than for the V$^0$ and V$^-$ for both Br and Cl vacancies. In Refs.~\cite{2000Stokbro, 2003Nakamura}, it was proposed that a positively charged vacancy attracts an electronegative adatom, which leads to a decrease in the diffusion barrier.

\begin{table}[h]
\caption{\label{NEB}Calculated activation energies of diffusion of vacancies in each of the charge states on the Si(100)-2$\times$1-Br and -Cl surfaces. Energies are given in electronvolts.}
\begin{ruledtabular}
\begin{tabular}{lccr}
Surface &  V$^+$ & V$^0$ & V$^-$ \\
\hline
    Si(100)-2$\times$1-Br & 0.22 & 0.88 & 1.14  \\
    Si(100)-2$\times$1-Cl & 0.44 & 1.14 & 1.33   \\
\end{tabular}
\end{ruledtabular}
\end{table}

To study the electronic structure of the Si(100)-2$\times$1-Br surface with a vacancy, we calculated the Crystal orbital Hamilton population (COHP) and Integrated Local Density of States (ILDOS) (Fig.~\ref{COHP}). Total charge distribution is given in the supplementary material. Figure~\ref{COHP}a shows the COHP for the neutral vacancy (the COHP plot is qualitatively the same for all three charge states of the vacancy). States with $-$COHP$>0$ ($-$COHP$<0$) are bonding (antibonding). The antibonding Si-Br orbital has a maximum in the range of 2--4\,eV, where it partially hybridizes with the antibonding orbital Si-Si$_{DB}$ of the dimer with a vacancy. In this range, the antibonding orbital between the Si$_{DB}$ atom of the dimer and the Si$_{ss}$ atom of the second layer has a smoother peak. According to the ILDOS plot, the antibonding orbital Si-Br has a predominantly $p_z$ component in the range of 2--4\,eV (Fig.~\ref{COHP}b). In the range from 0 to $-2$\,eV, the Si-Br bond has an antibonding orbital, whereas silicon bonds have a bonding orbital (Fig.~\ref{COHP}c), and in the range from $-2$ to $-4$\,eV all bonds have a bonding orbital (Fig.~\ref{COHP}d).

\begin{figure}[h]
 \includegraphics[width=\linewidth]{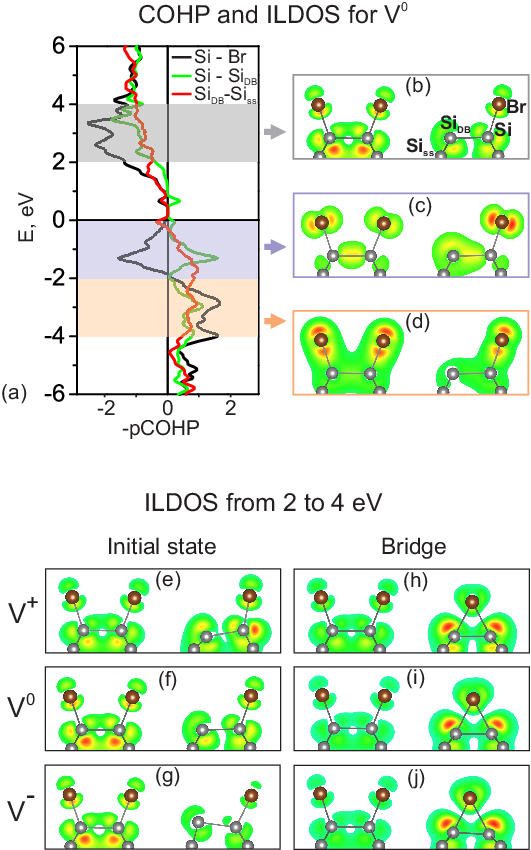}
\caption{\label{COHP}(a) Crystal orbital Hamilton population ($-$COHP) for the V$^0$ on Si(100)-2$\times$1-Br. Energy is given relative to the valence band maximum. Si denotes the surface atom with Br, Si$_{DB}$ is the Si surface atom holding the DB, and Si$_{ss}$ is the second layer atom. (b--j) Integrated Local Density of States (ILDOS) of the vacancy on Si(100)-2$\times$1-Br with isosurface level of 0.005\,e/{\AA}$^3$. (b--d) ILDOS of V$^0$ from 2 to 4\,eV (b), from $-2$ to 0\,eV (c), and from $-2$ to $-4$\,eV (d). ILDOS of V$^+$ (e,h), V$^0$ (f,i), and V$^-$ (g,j) is given at 2--4 eV in the initial (e--g) and bridge (h--j) states.}
\end{figure}

In the range of 2--4\,eV, corresponding to the $R(U_s$) peak at 2--4\,V, the COHP is highest at the Si-Br antibonding orbital, and the ILDOS for the V$^+$ is higher than for the V$^0$ and V$^-$ (Figs.~\ref{COHP}e--g). This is also evident in the empty state STM image, where a brighter halo around the V$^+$ corresponds to a higher density of unoccupied states (Fig.~\ref{figSTM}). The ILDOS of the antibonding orbital Si-Br differs for the three charge states of the vacancy only in the initial state, in which the diffusion process is initiated. In the bridge configuration, the electron density of the three charge states of a vacancy differs slightly (Figs.~\ref{COHP}h-j).

\section{Discussion}

The strong dependence of the hopping rate on the tunnel current (Fig.~\ref{RI}) clearly points to an electronic mechanism of diffusion. The field-induced mechanism cannot be the main mechanism, since in this case the hopping rate would not depend on the number of tunneling electrons \cite{2020Bohamud} and would exhibit a monotonic dependence on the bias voltage. Nevertheless, the electric field can also affect diffusion, for example, by changing the activation barrier. We also ruled out the mechanism of a direct chemical interaction between the tip and an adatom, as was ruled out for the H vacancy diffusion over the Si(100) surface \cite{2000Stokbro}. The contribution of thermal diffusion to the hopping rate can also be neglected, since we observed almost no diffusion at 77 K when scanning with a low current.

The hopping rate as a function of the positive sample bias has a broad peak at 2--4\,V (Fig.~\ref{RU}). In Ref.~\cite{2002Nakayama}, $R(U_s)$ also has a peak for bromine hopping, which was associated with $\sigma^{\ast}_{Si-Br}$ maxima in STS at the same voltage, and it was concluded that hops are caused by electron capture in the antibonding orbital. Note that the same mechanism was proposed for chlorine hopping \cite{2003Nakamura}, however, in contrast to our experiment and Ref.~\cite{2002Nakayama}, chlorine coverage was low (0.06 ML). Electrons with energies greater than the energy of the antibonding orbital can also be captured in $\sigma^{\ast}_{Si-Br}$ if they are first injected into the Si surface states \cite{2002Nakayama, 2003Nakamura}. The ejection of electrons through the surface states explains the broadening of the $R(U_s)$ peak and nonlocality of the diffusion process \cite{2001Quaade, 2010Bellec}. It should be stressed that in different studies, the STS peak was observed at different voltages, in particular, at $+0.8$\,V \cite{2002Nakayama} and $+2.0$\,V \cite{2006Nakayama} for Br and at $+1.2$\,V \cite{2003Nakamura}, $+1.4$\,V \cite{2002Nakayama}, and $+2.4$\,V \cite{2006Nakayama} for Cl. In our experiments, the broad peak of $R(U_s)$ is observed in the range 2--4\,V, which coincides with the maximum of the $p_z$ component of $\sigma^{\ast}_{Si-Br}$ in the calculations (Figs.~\ref{COHP}a, b). Thus, we also associate the broad peak of $R(U_s)$ with the capture of electrons in the $\sigma^{\ast}_{Si-Br}$ orbital either directly or through the surface states. It is noteworthy that the V$^+$ has a higher density of electronic states of $\sigma^{\ast}_{Si-Br}$ in the initial state compared to the V$^0$ and V$^-$, and hence, the V$^+$ has a higher probability of capturing an electron in $\sigma^{\ast}_{Si-Br}$.

We have obtained different dependences of the hopping rate on the current at the threshold voltage and at the peak voltage. At the threshold voltage of 2.5\,V, a power law $R \propto I_t^N$ with the exponent $N \approx 2$ suggests that two tunneling electrons are involved in the diffusion process. At the threshold voltage, the energy of one electron may be not sufficient to overcome the reaction barrier. A possible two-electron mechanism, that was proposed for Cl desorption \cite{2007Nakamura} is the localization of two electrons in the surface area near the vacancy, which destabilize the Si-Br bond. Another possible diffusion mechanism can be the excitation of vibrational modes by two tunneling electrons, as was proposed for H desorption from the Si(100) surface \cite{2003Soukiassian}.

At the peak voltage of 3.5\,V, the exponent $N \approx 1.5$, suggesting that one electron is sufficient to initiate hopping, although the two-electron process also operates. At 3.5\,V, an electron has enough energy to localize at the Si-Br antibonding orbital both directly or through the surface states. In previous studies, the injection of a single electron (or hole, at a negative voltage) was also proposed as a mechanism for the adatom diffusion on the Br- \cite{2002Nakayama}, Cl- \cite{2003Nakamura} (at a submonolayer coverage), and H-terminated \cite{2001Quaade, 2010Bellec} Si(100) surfaces. However, in Ref.~\cite{2002Nakayama} a one-electron mechanism was proposed for Br diffusion at a voltage ($+1.0$\,V) close to the threshold  ($+0.8$\,V). In contrast to our work, the experiments were (apparently) carried out at room temperature. This may suggest that thermal vibrations also contributed to diffusion at room temperature, or that the diffusion mechanisms differ in different ranges of parameters ($U_s = +1.0$\,V and $I_t \leqslant$ 0.05\,nA in Ref.~\cite{2002Nakayama}).

In contrast to other works \cite{2000Stokbro, 2002Nakayama, 2001Quaade, 2010Bellec, 2003Nakamura}, we studied the dependence of the diffusion rate on current and voltage for each vacancy separately and found that the $R(U_s)$ peak (Fig.~\ref{RU}) and the $R(I_t)$ slope (Fig.~\ref{RI}) do not coincide for different vacancies. In addition, we observed a change in the vacancy hopping rate with a change in the tip state, which is consistent with the previous study on the hydrogenated surface \cite{2001Quaade}. Consequently, different hopping rates of vacancies at the same scanning parameters can be related to the different tip states. However, the hopping rates for vacancies imaged on the same STM frame do not coincide (V$^+$1 and V$^+$2, V$^+$4 and V$^+$5), and hence hops are affected not only by the tip state. We assume that diffusion is also affected by the nonuniform distribution of defects and subsurface dopants. Indeed, band bending induced by surrounding impurities and defects leads to a shift in the position of the energy levels \cite{2015Labidi}. As a result, the charging voltage is different for DBs within the same STM image \cite{2015Labidi, 2022Pavlova}. Thus, we explain the slightly different hopping rates for various vacancies not only by different tip states, but also by nonidentical position of the $\sigma^{\ast}_{Si-Br}$ resonant orbital due to surrounding defects and impurities.

Unlike bromine vacancies, chlorine vacancies do not hop when scanning with a positive voltage. Our calculations agree with the experiment, indicating that the diffusion barrier for chlorine is much higher than for bromine (Table~\ref{NEB}). We believe that the main reason for the higher diffusion barrier for chlorine is the stronger Si--Cl interaction, as seen from the higher adsorption energy of Cl (4.27\,eV) compared to Br (3.59\,eV) \cite{2004Lee}. The stronger Si--Cl bond compared to Si--Br has also been pointed out in previous work as one of the reasons for the lack of diffusion on the chlorinated surface \cite{2002Nakayama}.

The stronger repulsive interaction between Br adatoms has been proposed as another reason for faster diffusion over a brominated surface than over a chlorinated one \cite{2002Nakayama}. The absence of bromine hopping on the Si(100) surface at a submonolayer coverage (during scanning at $U_s \leqslant +2$\,V and $I_t= 0.06$\,nA) was used as experimental proof of the role of the repulsive interaction \cite{2002Nakayama}. However, Cl hops at a low coverage and the hopping rate has a sharp peak at $+1.25$\,V (at $I_t= 0.8$\,nA) \cite{2003Nakamura}. Therefore, chlorine hops at a submonolayer coverage, but bromine does not, in contrast to the monolayer coverage, at which Br hops but Cl does not. This controversial behavior between halogen diffusion at a low and monolayer coverage cannot be explained by bond strength and repulsive interaction alone.

One of the possible explanations for this controversial behavior is the different charge of the DB at the Si atom located on the same dimer as the halogen atom. Indeed, the charge is critical, since only positively charged vacancies diffuse in our experiments. Our calculations also confirm that the activation barrier is much lower for both Cl and Br vacancy diffusion if the DB is positively charged, although an electronically stimulated reaction may not go along the path with the ground state, but turn into an excited one. Thereby, it can be assumed that DBs were neutral or negatively charged in experiments where the adatoms do not hop. Thus, in future studies of diffusion, it is very important to take into account the DB charge, which can be easily determined from the STM image. Note that the DB charge may vary depending on scanning parameters, doping level, and the presence of nearby impurities and defects.

\section{Conclusions}

We studied the vacancy diffusion on the Si(100)-2$\times$1-Br surface in an STM and demonstrated that the vacancy hopping strongly depends on the DB charge. We found that only positively charged vacancies hop when scanning at a positive voltage, while neutral and negatively charged vacancies do not hop, in agreement with our calculations of activation barriers for both Br and Cl. At the threshold voltage of 2.5\,V, we propose a two-electron diffusion mechanism, and at the peak voltage of 3.5\,V, the one-electron mechanism also operates by capturing an electron in the $\sigma^{\ast}_{Si-Br}$ orbital directly or through the surface states. For each vacancy, the diffusion rate has slightly different dependences on current and voltage, which we attribute to a different electrostatic environment due to nonuniform distribution of defects and subsurface dopants. The critical effect of the DB charge state on the halogen diffusion found in this work can also play a role in the case of other atoms and molecules diffusion over the Si(100) surface. Our finding is also important for fabrication of ordered  structures with charged DBs and should be considered when scanning such structures.

\section*{Supplementary Material}
See the supplementary material for the additional STM images showing the coexistence of  the V$^+$ and V$^0$, diffusion of  the V$^0$ and V$^-$, hopping rate as a function of the voltage for six different V$^+$s, and total charge distribution for the V$^+$, V$^0$, and V$^-$ in diffusion pathway.

\begin{acknowledgments}
This study was supported by the Russian Science Foundation under the grant No. 21-12-00299. We also thank the Joint Supercomputer Center of RAS for providing the computing power.
\end{acknowledgments}

\bibliography{paper_JCP_rev1}
\end{document}